\documentclass[9pt,twocolumn,twoside]{osajnl}
\usepackage{graphicx}
\usepackage{adjustbox}
\usepackage{graphicx}
\usepackage{color,xcolor}
\usepackage{color}
\usepackage{textcomp,gensymb}
\usepackage{upgreek}

\journal{ol} 

\setboolean{shortarticle}{true}

\title{Second harmonic and cascaded third harmonic generation in generalized quasi-periodic poled lithium niobate waveguides}

\author[1]{Li Zhang}
\author[1]{Xiao Wu}
\author[1,2]{Zhenzhong Hao}
\author[1]{Rui Ma}
\author[1]{Feng Gao}
\author[1,*]{Fang Bo}
\author[1,3]{Guoquan Zhang}
\author[1,4]{Jingjun Xu}
\affil[1]{The MOE Key Laboratory of Weak Light Nonlinear Photonics, TEDA Institute of Applied Physics and School of Physics, Nankai University, Tianjin 300457, China}
\affil[2]{haozhenzhong@mail.nankai.edu.cn}
\affil[3]{zhanggq@nankai.edu.cn}
\affil[4]{jjxu@nankai.edu.cn}
\affil[*]{Corresponding author: bofang@nankai.edu.cn}

\begin{abstract}
Lithium niobate (LN) thin film  has recently emerged as an important platform for nonlinear optical investigations for its large $\chi^{(2)}$ nonlinear coefficients and ability of light localization. In this paper, we report the first fabrication of LN on insulator (LNOI) ridge waveguides with generalized quasi-periodic poled superlattices using the electric field polarization technique and microfabrication techniques. Benefiting from the abundant reciprocal vectors, we observed efficient second-harmonic and cascaded third-harmonic signals in the same device, with the normalized conversion efficiency 1735\% W$^{-1}$cm$^{-2}$ and 0.41\% W$^{-2}$cm$^{-4}$, respectively. This work opens a new direction of nonlinear integrated photonics based on LN thin film.
\end{abstract}

\setboolean{displaycopyright}{true}

\begin{document}

\maketitle


Short-wavelength lasers are much more difficult to  achieve than the ones working at long wavelengths due to the shortage of suitable nonlinear materials and the serious material dispersion increasing rapidly towards higher frequencies. High-order harmonic generations, such as second-harmonic generation (SHG) and third-harmonic generation (THG), are important ways to realize short-wavelength continuous-wave lasers and pulse lasers with short-wavelength components. SHG is usually obtained by exploring the $\chi^{(2)}$ process in crystalline materials lacking central symmetry, such as LN \cite{Review-bo1, Review-bo2, Review-bo4, hzz, wg-Wang}. In contrast, THG can be achieved in any material based on the third-order nonlinear susceptibility $\chi^{(3)}$. Because $\chi^{(3)}$ is usually smaller than $\chi^{(2)}$ by orders of magnitude, an intense pump that may destroy a solid material is usually required to achieve an observable THG signal. Therefore, THG has usually been demonstrated under the pump of a focused high-power laser in gases and liquids \cite{THG-gas, THG-liquid}, which is easy to recover but difficult to integrate.

Cascaded second-order nonlinearity, the sum-frequency generation (SFG) of the fundamental wave and its SHG, is another method to achieve THG. For such a nonlinear process, the most critical issue is that the phase-matching conditions for the SHG and SFG must be satisfied simultaneously. Recently, SHG and cascaded SFG processes were demonstrated in LNOI photonic devices including microcavities \cite{OL-cavity, PRL-cavity} and waveguides \cite{waveguide-THG-PM, THG-chenxianfeng}. Unfortunately, high-order modes of the same/different polarization states are often involved in these cases, resulting in a relatively small mode overlap and a lower nonlinear coefficient, and thus reducing the nonlinear conversion efficiencies. A periodically poled crystal of multiple periods can overcome the above constraints by providing multiple reciprocal vectors to simultaneously fulfill the phase-matching conditions of the SHG and cascaded SFG processes leading to efficient THG. Furthermore, in a periodically poled optical waveguide\cite{dual-wg, FOS-zhu, TM-wg}, SHG and cascaded SFG with only fundamental modes get involved can be achieved to take advantage of the maximized mode overlap \cite{cavity-dual} as well as the most significant nonlinear coefficients, for example, $d_{33}$ of LN.

Initially, QPM works were mainly concentrated in two periodically poled crystals spliced together \cite{twoPPLN-THG2(2f-3f-3),twoPPLN-THG1}, accomplishing the SHG and SFG in segments. However, this approach shortens the nonlinear interaction length to a certain extent by separating the coupled length into a few parts. Ming et al. first experimentally introduced the Fibonacci optical superlattice into a bulk LiTaO$_{3}$, which achieved QPM processes for SHG and THG simultaneously \cite{FOS-zhu}. They further extended it to the generalized quasi-periodic superlattice (GQPS) \cite{Projection1}. GQPS is more universal, and not restricted to specific wavelength or geometric construction. In addition, compared with the two-step scheme, the THG conversion efficiency of a GQPS is increased by a factor of 2.37 with the same waveguide length involved.

Optical waveguides offer more efficient nonlinear optical process than the bulk crystal benefiting from its transverse light confinement. LN waveguides fabricated with titanium diffusion and reverse-proton exchange \cite{Ti-LN, RPE-LN} show better nonlinear performance than bulk LN devices, although they have a relatively large mode area (with a diameter of about 10 $\upmu$m and a small core-to-cladding index contrast ($\Delta n \sim 0.02$). In contrast, LNOI ridge waveguides with a $\sim$1 $\upmu$m$^{2}$ mode area provide much tight optical confinement enabling more efficient wavelength conversion processes. Here, we introduce GQPS in the LNOI ridge waveguide, featuring tight light confinement in space and giant second-order nonlinear coefficient. The ridge waveguide was fabricated on LNOI wafer with microfabrication and electrically poling techniques. QPMs for SHG and the cascaded SFG were realized simultaneously on one chip. Taking advantage of the large spatial mode overlap among fundamental waveguide modes and  maximum nonlinear coefficient $d_{33}$ of LN, SHG and THG with conversion efficiencies of 1735\% W$^{-1}$cm$^{-2}$ and 0.41\% W$^{-2}$cm$^{-4}$ were observed, respectively. 

\section{QPMs of SHG and the cascaded THG in GQPS LN waveguides}

\begin{figure}[t]
\centering
\fbox{\includegraphics[width=0.95 \linewidth]{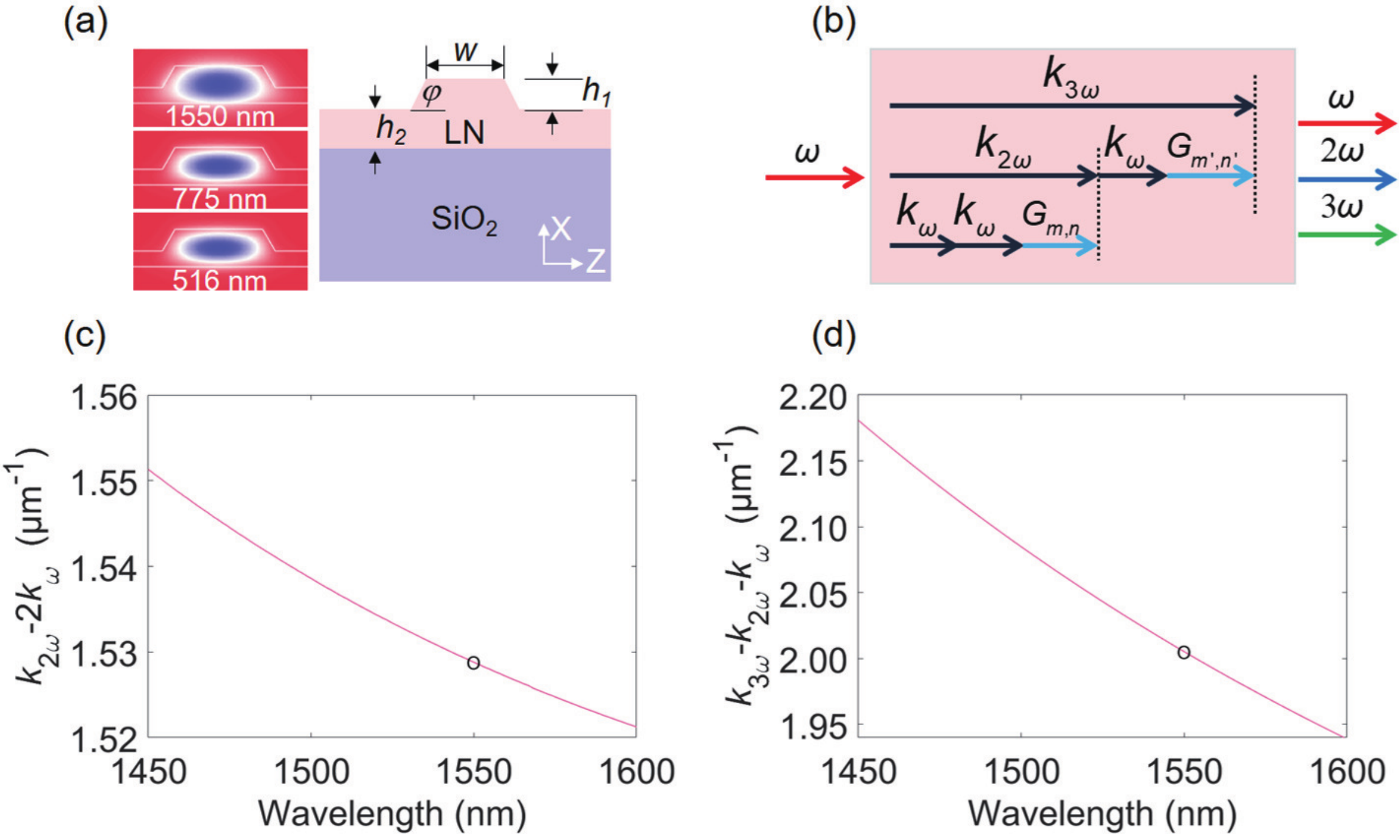}}
\caption{(a) Schematic of an x-cut LN waveguide. Finite element method (FEM) simulations of the profiles for fundamental modes at 1550 nm, 775 nm, and 516 nm, respectively. (b) Schematic diagram of the process of THG in a GQPS LN. The reciprocal vectors of the GQPS, ${G}_{{m},{n}}$ and $G_{m^{\prime}, n^{\prime}}$ are used to compensate for the mismatch of wave vectors in the SHG and SHG processes, respectively. (c, d) show the dependence of the original wavevector mismatches, $k_{2\omega}-2 k_{\omega}$ and $k_{3\omega}-k_{2\omega}-k_{\omega}$, on wavelengths, respectively. The data points marked by circles are used in subsequent experiments.}
\label{fig1}
\end{figure}

Consider SHG and the cascaded SFG processes in a ridge LN waveguide with a superlattice domain structure. The cross-section of the waveguide is defined by the width $w$ of the top surface, the tilt angle $\varphi$ of the side wall, the etching depth $h_1$, and the thickness  $h_2$ of the slab, as labeled in Fig. \ref{fig1}(a). We assume only fundamental modes indicated in Fig. \ref{fig1}(a) get involved in the nonlinear processes to get efficient wavelength conversion. As shown in Fig. \ref{fig1}(b), the reciprocal vectors $G_{m,n}$ and $G_{m\prime,n\prime}$ are employed to fulfill QPM conditions for the SHG and cascaded SFG processes, respectively. The final mismatches in wavevector for the SHG and SFG processes after compensation are written as $\Delta k_{\rm{SHG}}=k_{2\omega}-2k_{\omega}-G_{m, n}$, and $\Delta k_{\rm{SFG}}=k_{3\omega}-k_{2\omega}-k_{\omega}-G_{m^{\prime}, n^{\prime}}$, 
respectively. Here $k_{\omega}$, $k_{2\omega}$, $k_{3\omega}$ are the wave vector of the pump, SHG and THG, respectively. We calculated the original wavevector mismatches, $k_{2\omega}-2k_{\omega}$ and $k_{3\omega}-k_{2\omega}-k_{\omega}$, in a given LN waveguide at different wavelengths and temperature. The original wavevector mismatch due to dispersion determines the reciprocal vectors that are required to achieve QPM. As shown in Fig. \ref{fig1}(c, d), the required reciprocal vectors decrease with the redshift of the pump wavelength.  In simulation, we set $w$, $\varphi$, $h_1$, and $h_2$ to be 1.42 $\upmu$m, 60$^{\circ}$, 350 nm, and 250 nm, respectively. These parameters are the same as those 
of the fabricated waveguides.

\begin{figure}[!hbt]
\centering
\fbox{\includegraphics[width=0.95 \linewidth]{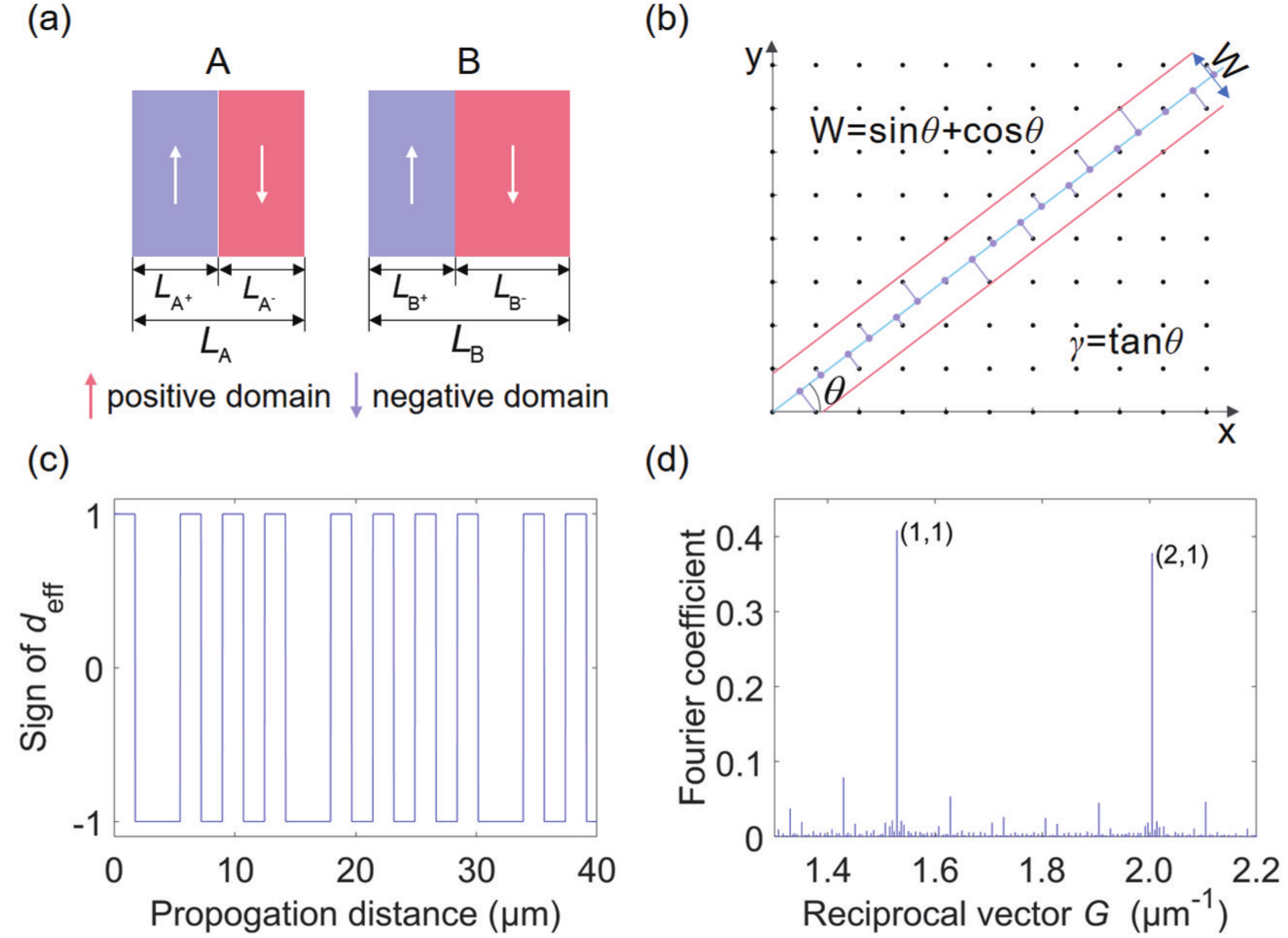}}
\caption{(a) Building blocks A and B of GQPS, each composed of one positive and one negative domain.
(b) Sequence arrangement of a GQPS by the projection method \cite{Projection1, Projection2}. (c) Schematic diagram showing part of the GQPS indicated by the sign of $d_{\rm eff}$ in this work. (d) Reciprocal vectors corresponding to the domain structure in (c).}
\label{fig2}
\end{figure}

In a single periodically poled LN (SPPLN), all the supported reciprocal vectors are odd multiples of the fundamental reciprocal vector $2\pi/\Lambda$. The initial phase mismatches due to dispersion for SHG and SFG tend to be unequal in most cases. Therefore, it is challenging to utilize an SPPLN waveguide to access QPMs for cascaded second-order nonlinear processes with light waves of significantly different wavelengths unless a specific waveguide structure is carefully designed to make the original wave vector mismatch equal. In contrast, 1D GQPS with two or more linearly independent structural parameters  \cite{FOS-zhu, FOS-3} not only supports more reciprocal vectors but also provides greater structural design flexibility by introducing extra degrees of freedom. 

Here, we discuss the case of linearly independent structures consisting of two building blocks, labeled as A and B, according to a particular arrangement rule \cite{Projection1, Projection2}. Each block is composed of a positive and a negative domain with inversed polarization, as shown in Fig. \ref{fig2}(a). For simplicity, we assume the positive domain has the same length ($L_{\rm{A^{+}}}=L_{\rm{B^{+}}}=l$), while the negative domain has a different length ($L_{\rm{A^{-}}}\neq L_{\rm{B^{-}}}$). Apparently, when $L_{\rm{A^{+}}}=L_{\rm{B^{+}}}=L_{\rm{A^{-}}}=L_{\rm{B^{-}}}$, the GQPS structure reduces to an SPPLN. The reciprocal vectors provided by a 1D GQPS can be written as \cite{FOS-FFT} ${G}_{{m},{n}}=\frac{2 \pi({m}+{n} \gamma)}{{D}}$,
where $\gamma$ is an arbitrary number and ${D}=\gamma{L}_{\rm{{A}}}+{L}_{\rm{{B}}}$ is the average structure parameter. ${G}_{{m},{n}}$ depend on two independent structure parameters $\gamma$ and ${D}$, and are indexed by two integers $m$ and $n$. They are different from the reciprocal vectors in an SPPLN determined by one structure parameter and labeled by one integer. The Fourier coefficients are $g_{m, n}=2(1+\gamma) l / D \operatorname{sinc}\left(G_{m, n} l / 2\right) \operatorname{sinc}\left(X_{m, n}\right)$ where $X_{m, n}=\pi (1+\gamma)\left(m L_{\mathrm{A}}-n L_{\mathrm{B}}\right) / D$. 
The corresponding effective nonlinear coefficient is $d_{{\rm{eff}},(m, n)}={g}_{{m},{n}}d_{33}$. When the geometrical structure of the waveguide is determined as shown in Fig. \ref{fig1}(a), we can obtain the required reciprocal vectors as 1.53 $\upmu$m$^{-1}$ for SHG and 2.00 $\upmu$m$^{-1}$ for SFG, respectively. In order to get higher Fourier coefficients, by selecting $m$=1, $n$=1, $m^{\prime}$=2, $n^{\prime}$=1 , we can uniquely determine $\gamma$=2.21, ${D}$=13.19 $\upmu$m. The arrangement sequence can be obtained according to the projection theory\cite{Projection1, Projection2} illustrated in Fig. \ref{fig2}(b). Here, the projection sequence we use is BAABAAABA$\cdots$. The sequence arrangement is the Fibonacci sequence \cite{FOS-zhu}, when $\gamma=\tau=(1+\sqrt{5})/2$ that is golden ratio. By using a simulated annealing algorithm, the structure of the positive and negative domains of blocks A and B are optimized to maximize the product of Fourier coefficients of the reciprocal vectors used in nonlinear processes. Figure \ref{fig2}(c) shows the variation of $d_{\rm eff}$ corresponding to the partial sequence of the GQPS waveguide. Here, $L_{\rm{A}}$=3.49 $\upmu$m, $L_{\rm{B}}$=5.47 $\upmu$m, and $L_{\rm{{A^{+}}}}=L_{\rm{{B^{+}}}}$=1.74 $\upmu$m. The effective nonlinear coefficients associated with $G_{1,1}$ and $G_{2,1}$ are 0.41$d_{33}$, 0.38$d_{33}$, respectively, as shown in Fig. \ref{fig2}(d). Besides the nonlinear coefficients, the SHG/SFG efficiency depends on the 
 effective area and spatial overlap of the interacting waveguide modes.

\section{Preparation and characterization of GQPS LN waveguides}


\begin{figure}[t]
\centering
\fbox{\includegraphics[width=0.95 \linewidth]{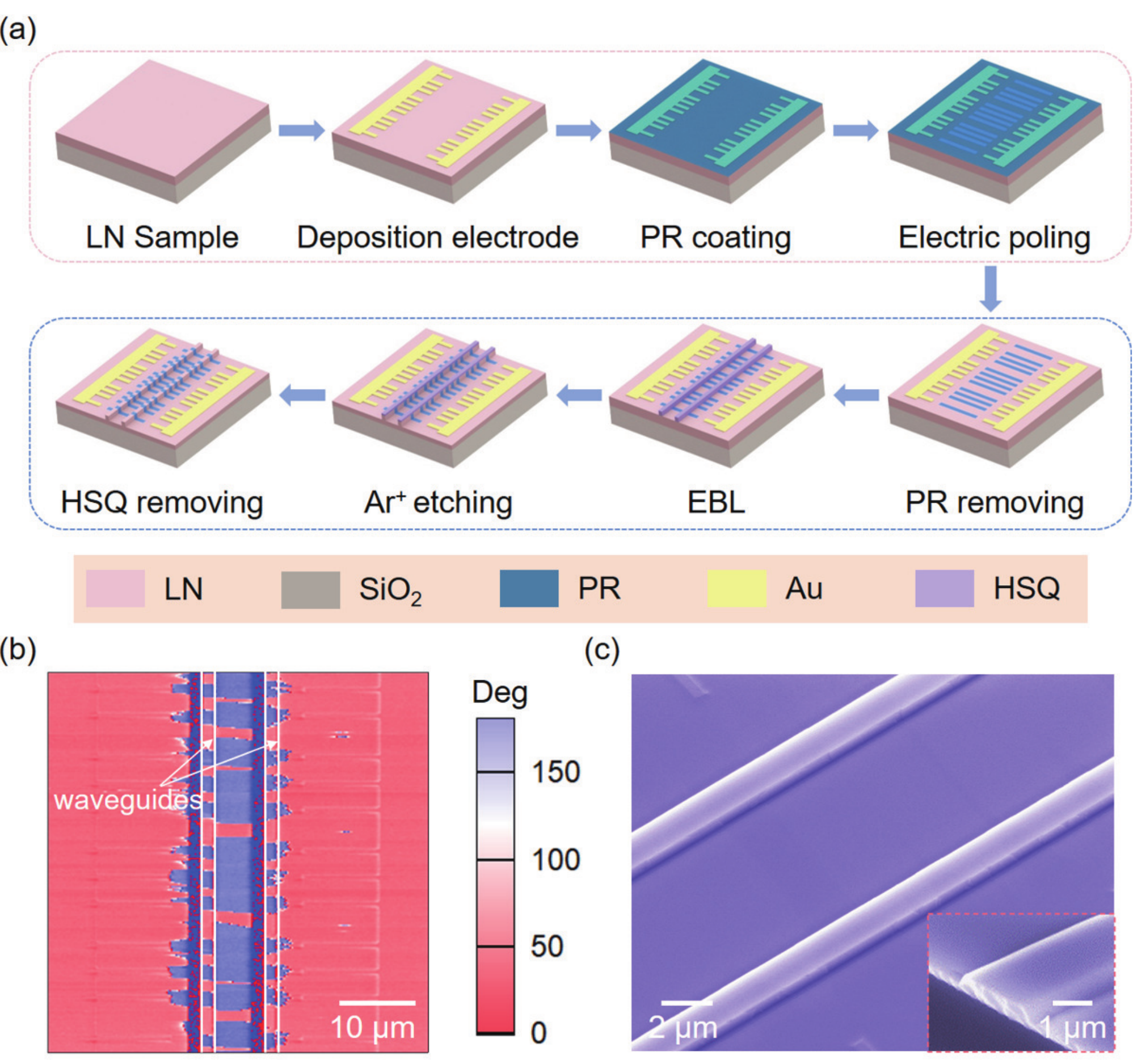}}
\caption{(a) Schematic diagram of the preparation process of GQPS LN waveguides. PR: photoresist, EBL: electron beam lithography, HSQ: hydrogen silsesquioxane.  (b) PFM image showing the domain structure of the LN waveguides. (c) SEM  image of the GQPS LN waveguides. Inset: waveguide facet.}
\label{fig3}
\end{figure}

The GQPS LN waveguides were fabricated on a 600 nm x-cut LNOI wafer from NANOLN. The preparation recipe roughly consists of two parts: electrically poling of the LN thin film and the fabrication of ridge waveguides, as indicated by the dashed boxes in Fig. \ref{fig3}(a). A detailed description of the preparation process can be referred to our previous paper \cite{wx-wg}. The domain structures of the LN waveguide were characterized by a piezoelectric force microscope (PFM). The PFM image of typical GQPS LN waveguides circled by white lines is shown in Fig. \ref{fig3}(b). It can be seen that the adjacent domains have a 180$^{\circ}$ phase difference, where the reversed and intact regions are marked by blue and red colors, respectively. The average length of the positive domain along the waveguide direction is 1.83 $\upmu$m, which is 90 nm longer than the design. Such a deviation can be optimized by reducing the polarization time and adjusting the electrode structure. In order to get good experimental results avoiding machining errors, we placed two waveguides between the poling electrodes. Figure \ref{fig3}(c) and its inset present the scanning electron microscopy (SEM) image of the GQPS LN waveguides and facet. The geometric fluctuation could be vaguely seen on the side wall mainly attributed to the different etching rates of ±z axis when the hydrofluoric acid removes the HSQ residual mask. That may cause additional transmission loss. The waveguide propagation loss for the pump in 7-mm waveguide with 4-mm PPLN region was measured to be 5.0 dB/cm by using the Fabry-Perot interference method \cite{FP-loss}. The fiber-chip coupling losses of SHG and THG were obtained through the finite-difference time-domain simulation. The validity of the theoretical model was verified by comparing the simulated and the measured coupling losses of the pump beam. Combining the measured fiber-fiber coupling losses and the simulated fiber-chip coupling losses of SHG and THG, we further estimated the waveguide propagation losses for the SHG and THG are 9.8 dB/cm and 15.0 dB/cm, respectively. 


\section{SHG and THG in GQPS LN waveguides}



\begin{figure}[t]
\centering
\fbox{\includegraphics[width=0.95\linewidth]{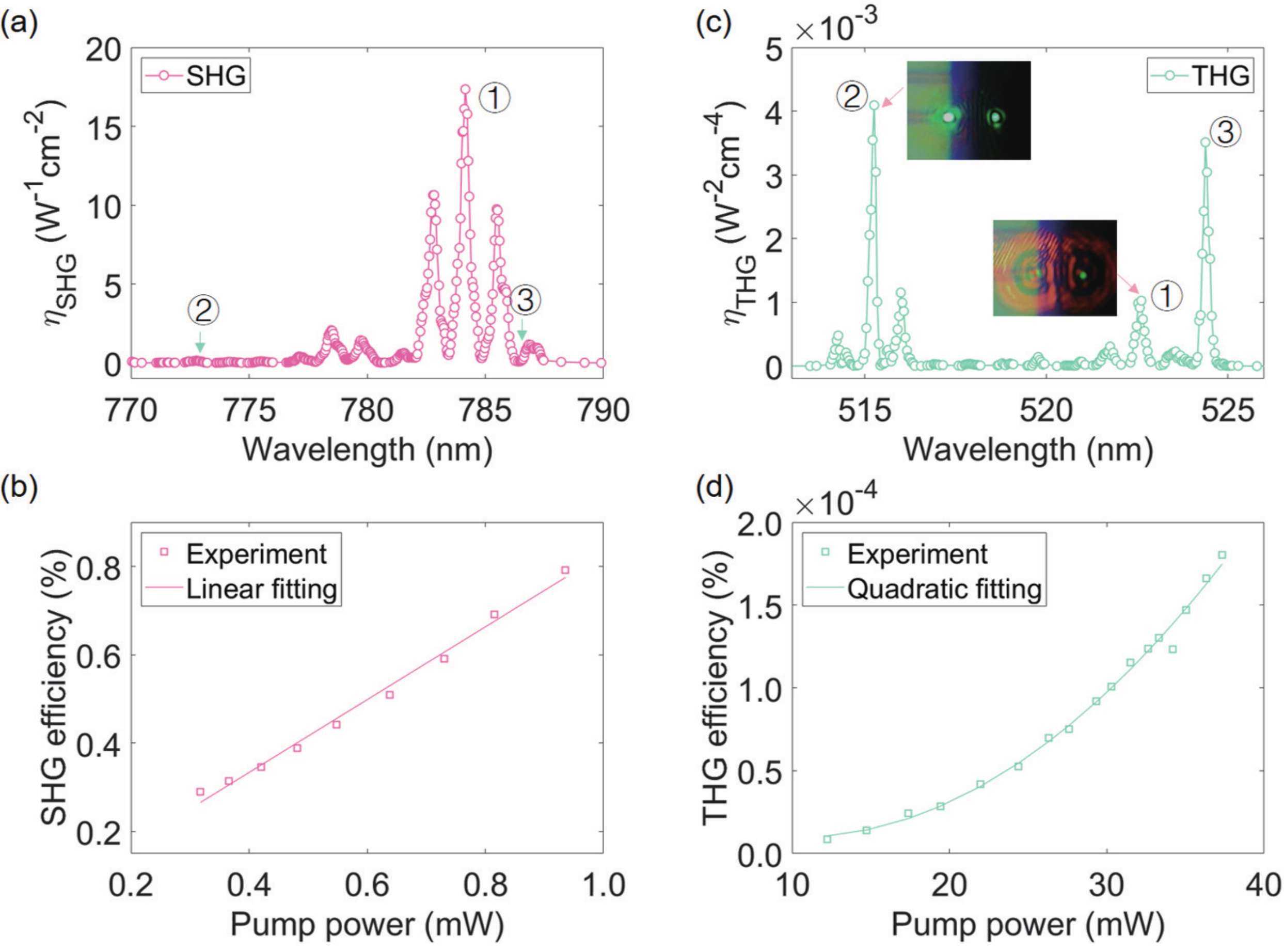}}
\caption{ (a, c) show the dependence of SHG and THG on the pump wavelength, respectively. Inset of (c) shows the scattered SHG and THG light at the output waveguide facet detected by a CCD camera. (b, d) SHG$_{1}$ and THG$_{2}$ efficiency as a function of pump power.}
\label{fig4}
\end{figure}

The experimental setup similar to that we used in our previous work \cite{wx-wg} was utilized to characterize the cascaded harmonic generation of the prepared GQPS waveguide. A series of SHG signals were obtained by adjusting the pump polarization and scanning the pump wavelength. Figure \ref{fig4}(a) shows the dependence of normalized conversion efficiency on the pump wavelength with 1.1 mW on-chip pump power. We collected 0.96 $\upmu$W SHG signal with 3.72 mW incident light power at the pump wavelength of 1568.2 nm labeled as $\mathrm{SHG_{1}}$, corresponding to a 1735\% W$^{-1}$cm$^{-2}$ on-chip conversion efficiency. The relatively high side lobes are probably attributed to the imperfect domain structures, and the inhomogeneous thickness of the waveguide over its entire length \cite{side-lobe}. The dependence of the SHG$_{1}$ efficiency with the pump power is shown in Fig.\ref{fig4}(b); the linear fitting indicates a conversion efficiency of 833\%  W$^{-1}$.

At lower power, the THG detected by the spectrometer was not obvious. Therefore, in order to get clear THG signals, we further increased the on-chip power to 26.75 mW. Figure \ref{fig4}(c) shows the dependence of the normalized THG conversion efficiency on the pump wavelength. When the SHG power reached its highest, the normalized conversion efficiency of the  $\mathrm{THG_{1}}$ was measured to be 0.1\%  W$^{-2}$cm$^{-4}$ with a 0.32-nm full width half maximum. It can be clearly seen that the THG signals at positions 2 and 3 are stronger than 1 in Fig. \ref{fig4}(c), while the $\mathrm{SHG_{2,3}}$ is relatively weak. This is also confirmed by the intensity of scattered harmonic light at the output waveguide facet captured by a CCD camera in the insets of Fig. \ref{fig4}(c). It is mainly caused by the uneven waveguide geometry and imperfect domain quality \cite{wg-chen-THG}. The normalized efficiencies for $\mathrm{THG_{2}}$ and $\mathrm{THG_{3}}$ are 0.41\% W$^{-2}$cm$^{-4}$ and 0.35\% W$^{-2}$cm$^{-4}$, while the normalized efficiency for $\mathrm{SHG_{2}}$ and $\mathrm{SHG_{3}}$ are 10.51\% W$^{-1}$cm$^{-2}$ and 23.73\% W$^{-1}$cm$^{-2}$ as shown in Fig.\ref{fig4}(a), respectively. The efficiency of the $\mathrm{THG_{2}}$ varying with the pump power is shown in Fig. \ref{fig4}(d), exhibiting a quadratic relationship. The appearance of the multiple THG signals is attributed to the abundant reciprocal vectors provided by the generalized quasi-periodic LN waveguide. Among them, the average pump, SHG and THG fiber-chip coupling losses are 5.26 dB/facet, 10.40 dB/facet, and 11.87 dB/facet without edge polishing.


To verify the experimental is consistent with theoretical results, we use FEM to theoretically calculate the field distribution and mode effective refractive index of the fundamental, SHG and THG modes. Numerical simulation shows that the normalized THG conversion efficiencies of TE fundamental modes are 2799\% W$^{-2}$cm$^{-4}$ with the waveguide transmission loss taking into account \cite{THG-efficiencybook, loss-efficiency}. Considering the measured THG efficiency is nearly five orders of magnitude lower than the theoretical conversion efficiency, we also calculated the conversion efficiency with high order mode participating in the nonlinear processes. Assuming that the modes in 1550-nm, 780-nm and 520-nm bands are TE$_{0}$, TE$_{0}$, TE$_{2}$, respectively, we can obtain a THG conversion efficiency of  74\% W$^{-2}$cm$^{-4}$. In this case, the effective nonlinear susceptibility, the effective mode area, and the spatial mode overlap factor for SFG were calculated to be 0.38, 1.47 $\upmu$m$^2$, and 0.22, respectively. Therefore, we have reason to guess that, reciprocal vectors with a lower Fourier coefficient or high-order waveguide modes participate in the nonlinear processes, which leads to the THG conversion efficiency degradation. The THG efficiency can be further improved by optimizing the domain structure and uniformity of the LN waveguide. For an SPPLN, the effective nonlinear coefficient is $2 d_{33} / \pi$, which corresponds to the maximum conversion efficiency $\eta_{{_{\rm{SHG}}}}\sim$ 4500\% W$^{-1}$cm$^{-2}$. It is inevitable to sacrifice a few effective nonlinear coefficients of SHG to generate THG in the same waveguide.
\section{Summary}
In short, we prepared a generalized quasi-periodic LN ridge waveguide on a chip for the first time, based on which the cascaded second-order parametric process was realized. The measured SHG and THG conversion efficiencies are 1735$\%$ W$^{-1}$cm$^{-2}$ and 0.41$\%$ W$^{-2}$cm$^{-4}$, respectively. Cascading second-order nonlinear optical effects may find applications in many fields, such as the generation of multi-color solitons, higher harmonics, and entangled/squeezed photon states. 








\section{Funding Information}

This work was supported by the National Key Research and Development Program of China (Grant No. 2019YFA0705000), the National Natural Science Foundation of China (Grant Nos. 12034010, 92050111, 12134007, 11734009, 92050114, 12004197, and 12074199), the Higher Education Discipline Innovation Project (Grant No. B07013).

\section{Disclosures}

The authors declare no conflicts of interest.










\bibliography{GQPPLN}

\bibliographyfullrefs{GQPPLN}


\ifthenelse{\equal{\journalref}{aop}}{%
\section*{Author Biographies}
\begingroup
\setlength\intextsep{0pt}
\begin{minipage}[t][6.3cm][t]{1.0\textwidth} 
  \begin{wrapfigure}{L}{0.25\textwidth}
    \includegraphics[width=0.25\textwidth]{john_smith.eps}
  \end{wrapfigure}
  \noindent
  {\bfseries John Smith} received his BSc (Mathematics) in 2000 from The University of Maryland. His research interests include lasers and optics.
\end{minipage}
\begin{minipage}{1.0\textwidth}
  \begin{wrapfigure}{L}{0.25\textwidth}
    \includegraphics[width=0.25\textwidth]{alice_smith.eps}
  \end{wrapfigure}
  \noindent
  {\bfseries Alice Smith} also received her BSc (Mathematics) in 2000 from The University of Maryland. Her research interests also include lasers and optics.
\end{minipage}
\endgroup
}{}

\end{document}